%% file: main.tex
\documentclass[12pt]{article}
\usepackage[title]{appendix}
\usepackage{rotating, lscape}
\usepackage{float}
\usepackage{graphicx, color}
\usepackage{threeparttable, booktabs, longtable, threeparttablex, tabularx, multirow, pdflscape}
\usepackage{placeins}

\usepackage[natbibapa]{apacite}

\usepackage{courier}
\usepackage[utf8]{inputenc}
\usepackage[T1]{fontenc}

\usepackage{amssymb,amsmath,mathrsfs,amsthm, bbm}
\usepackage{accents}

\usepackage{microtype}
\usepackage[colorlinks=true, citecolor=black, urlcolor=blue]{hyperref}
\usepackage{afterpage}
\usepackage{caption}
\usepackage{amsmath}
\usepackage{subfigure}
\usepackage{setspace}
\usepackage{graphicx}

\RequirePackage[font=small,format=plain,labelfont=bf,textfont=it]{caption}
\usepackage{chngcntr}
\usepackage{threeparttable}
\usepackage{amsthm}
\usepackage{ragged2e}
\usepackage{subcaption}

\newtheorem{assumption}{Assumption}

\usepackage{textcomp}
\usepackage{tkz-graph}
\usetikzlibrary{shapes.geometric}

\usepackage{xcolor}
\hypersetup{
  colorlinks   = true, 
  urlcolor     = blue, 
  linkcolor    = blue, 
  citecolor   = blue 
}

\begin{document}

\begin{titlepage}
\title{Testing Effect Homogeneity and Confounding\\in High-Dimensional Experimental and Observational Studies}
\author{Ana Armendariz\hspace{.2cm}\\
    \small{University of St.\ Gallen, School of Economics and Political Science}\vspace{.2cm}\\ 
    Martin Huber\hspace{.2cm}\\
    \small{University of Fribourg, Dept.\ of Economics}\vspace{.2cm}\\}
\date{\today}
\maketitle

\begin{abstract}

We propose a framework for testing the homogeneity of conditional average treatment effects (CATEs) across multiple experimental and observational studies. Our approach leverages multiple randomized trials to assess whether treatment effects vary with unobserved heterogeneity that differs across trials: if CATEs are homogeneous, this indicates the absence of interactions between treatment and unobservables in the mean effect. Comparing CATEs between experimental and observational data further allows evaluation of potential confounding: if the estimands coincide, there is no unobserved confounding; if they differ, deviations may arise from unobserved confounding, effect heterogeneity, or both. We extend the framework to settings with alternative identification strategies, namely instrumental variable settings and panel data with parallel trends assumptions based on differences in differences, where effects are identified only locally for subpopulations such as compliers or treated units. In these contexts, testing homogeneity is useful for assessing whether local effects can be extrapolated to the total population. We suggest a test based on double machine learning that accommodates high-dimensional covariates in a data-driven way and investigate its finite-sample performance through a simulation study. Finally, we apply the test to the International Stroke Trial (IST), a large multi-country randomized controlled trial in patients with acute ischaemic stroke that evaluated whether early treatment with aspirin altered subsequent clinical outcomes. Our methodology provides a flexible tool for both validating identification assumptions and understanding the generalizability of estimated treatment effects.
\noindent \\
\vspace{0in}
\vspace{0in}

\bigskip

\renewcommand{\thefootnote}{\fnsymbol{footnote}}
\footnotetext[1]{We thank Federica Mascolo for helpful comments.}
\renewcommand{\thefootnote}{\arabic{footnote}}

\end{abstract}

\vfill

\setcounter{page}{0}
\thispagestyle{empty}

\end{titlepage}
\pagebreak \newpage

\singlespacing

\input{intro}
\input{literature}
\input{method}

\input{simulation}

\input{data}

\input{discussion}

\singlespacing
\setlength\bibsep{0pt}

\clearpage

\FloatBarrier

\addcontentsline{toc}{section}{References}
\bibliographystyle{apacite}
\bibliography{references}

\section*{Appendix } \label{sec:appendixa}
\setcounter{equation}{0} 
\renewcommand{\theequation}{A.\arabic{equation}}

\addcontentsline{toc}{section}{Appendix}

\subsection*{Proof: Moment condition and Neyman orthogonality for $\psi^{\mathrm{CLATE}}$}

\paragraph{Step 1: Definitions.}
Let
\[
m_{w,x,z}=E[Y\mid W=w,X=x,Z=z],\quad 
r_{w,x,z}=E[D\mid W=w,X=x,Z=z],
\]
\[
\bar g_{X,z}=m_{1,X,z}-m_{0,X,z},\qquad 
\bar h_{X,z}=r_{1,X,z}-r_{0,X,z}.
\]
Define the augmentation (or debiasing) terms containing the residuals that are weighted by the inverse of the propensity score:
\begin{align*}
A^{(Y)}_{X,z}&=
\frac{(Y-m_{1,X,z})\mathbf{1}\{W=1,Z=z\}}{\pi_{1,z}(X)}
-\frac{(Y-m_{0,X,z})\mathbf{1}\{W=0,Z=z\}}{\pi_{0,z}(X)}\\
A^{(Y)}_{X,z^-}&= \frac{(Y - m_{1,X,z^-})\mathbf{1}\{W=1,Z\neq z\}}{\pi_{1,z^-}(X)}
- \frac{(Y - m_{0,X,z^-})\mathbf{1}\{W=0,Z\neq z\}}{\pi_{0,z^-}(X)}.
\end{align*}
and $A^{(D)}_{X,z}, A^{(D)}_{X,z^-}$ analogously with $D$ instead of $Y$.
Let
\[
g_{X,z}=\bar g_{X,z}+A^{(Y)}_{X,z}, \qquad
h_{X,z}=\bar h_{X,z}+A^{(D)}_{X,z}.
\]
The  DR cross-products are
\[
\Theta_{X,z}=g_{X,z}h_{X,z^-}-g_{X,z^-}h_{X,z},
\]
which composed of the  plain regression cross-products
\[
\bar\Theta_{X,z}=\bar g_{X,z}\bar h_{X,z^-}-\bar g_{X,z^-}\bar h_{X,z},
\qquad
\]
and the augmentation term cross-products
\[
\mathcal{A}^{(Y)}_{X,z}= A^{(Y)}_{X,z}h_{X,z^-}-A^{(Y)}_{X,z^-}h_{X,z},\quad \mathcal{A}^{(D)}_{X,z}= A^{(D)}_{X,z}g_{X,z^-}-A^{(D)}_{X,z^-}g_{X,z}.
\]
The difference in the expectation of score function \eqref{scoreCLATE} and $\theta$ corresponds to the map
\begin{align}\label{mfunc}
M(\eta)=E\Big[\sum_{z=1}^{L}\Big\{\bar\Theta_{X,z}^{2}
+\Theta_{X,z}
+2\bar\Theta_{X,z}\mathcal{A}^{(Y)}_{X,z}
+2\bar\Theta_{X,z}\mathcal{A}^{(D)}_{X,z}\Big\}\Big].
\end{align}
We note that quadratic augmentation terms like $(\mathcal{A}^{(Y)}_{X,z})^{2}$ or $(\mathcal{A}^{(D)}_{X,z})^{2}$ could be added in the expectation defining $M(\eta)$, which would recognize that the null hypothesis in \eqref{hypothesisCLATE} is based on $E\!\left[\sum_{z=1}^L \Big(\Theta_{X,z}^2 + \Theta_{X,z} \Big)\right]$. However, such terms are of second order in the residuals and for this reason, they do not affect Neyman orthogonality, and are not required for defining the moment condition underlying our test either. They are for this reason not included in $M(\eta)$ and the following proof.

\paragraph{Step 2: Moment condition.}

Expanding $\Theta_{X,z}$ yields
\begin{align}\label{elements}
\Theta_{X,z}
&=(\bar g_{X,z}+A^{(Y)}_{X,z})(\bar h_{X,z^-}+A^{(D)}_{X,z^-})
-(\bar g_{X,z^-}+A^{(Y)}_{X,z^-})(\bar h_{X,z}+A^{(D)}_{X,z})\\
&=\underbrace{\bar g_{X,z}\bar h_{X,z^-}}_{(a)}
+\underbrace{\bar g_{X,z}A^{(D)}_{X,z^-}}_{(b)}
+\underbrace{A^{(Y)}_{X,z}\bar h_{X,z^-}}_{(c)}
+\underbrace{A^{(Y)}_{X,z}A^{(D)}_{X,z^-}}_{(d)}\notag\\
&\quad
-\underbrace{\bar g_{X,z^-}\bar h_{X,z}}_{(e)}
-\underbrace{\bar g_{X,z^-}A^{(D)}_{X,z}}_{(f)}
-\underbrace{A^{(Y)}_{X,z^-}\bar h_{X,z}}_{(g)}
-\underbrace{A^{(Y)}_{X,z^-}A^{(D)}_{X,z}}_{(h)}.\notag
\end{align}
Grouping, we see that
\[
\Theta_{X,z}=\bar\Theta_{X,z}
+\Big[(b)+(c)-(f)-(g)\Big]
+\Big[(d)-(h)\Big].
\]
We note that score function \eqref{scoreCLATE} is equal to 
\begin{align}\label{scorefunctappendix}
\psi^{\mathrm{CLATE}}(O,\theta,\eta)
=\sum_{z=1}^{L}\Big\{\bar\Theta_{X,z}^{2}
+\Theta_{X,z}
+2\bar\Theta_{X,z}\mathcal{A}^{(Y)}_{X,z}
+2\bar\Theta_{X,z}\mathcal{A}^{(D)}_{X,z}\Big\}-\theta.
\end{align}
Therefore,
\[
M(\eta)=E[\psi^{\mathrm{CLATE}}(W,X,Z;\eta)+\theta].
\]
At the true nuisance functions $\eta_0$, the augmentation terms conditionally mean zero given $X$:
\[
E[A^{(Y)}_{X,z}\mid X]=E[A^{(D)}_{X,z}\mid X]=0.
\]
This implies that terms $(b)$, $(c)$, $(f)$, and $(g)$ in equation \eqref{elements}, which involve augmentation terms, are equal to zero. Also terms $(d)$ and $(h)$, which are products of augmentation terms, are conditionally mean zero. It follows that $E[\Theta_{X,z}|X]=E[\bar \Theta_{X,z}|X]$. Furthermore, the terms $2\bar\Theta_{X,z}\mathcal{A}^{(Y)}_{X,z}$ and 
$2\bar\Theta_{X,z}\mathcal{A}^{(D)}_{X,z}$ in equation \eqref{scorefunctappendix} are conditionally mean zero as well. It follows by the law of iterated expectations that
\begin{align}
E\Big[\sum_{z=1}^{L}\big\{\bar\Theta_{X,z}^{2}
+\Theta_{X,z}
+2\bar\Theta_{X,z}\mathcal{A}^{(Y)}_{X,z}
+2\bar\Theta_{X,z}\mathcal{A}^{(D)}_{X,z}\big\}\Big]
&=E\Big[\sum_{z=1}^{L}\bar\Theta_{X,z}^{2}
+\bar\Theta_{X,z}\Big].
\end{align}
As the term $E\Big[\sum_{z=1}^{L}\bar\Theta_{X,z}^{2}
+\bar\Theta_{X,z}\Big]$ corresponds to the definition of $\theta_0$ in equation \eqref{hypothesisCLATE}, the moment condition
\begin{align}
E[\psi^{\mathrm{CLATE}}(O,\theta_0,\eta)]=0
\end{align}
holds, implying that the expectation of the score function is zero at the true values of the nuisance parameters $\eta$ and the test statistic $\theta_0$.

\subsubsection*{Step 3: Neyman orthogonality}


\paragraph{\textit{Perturbations in outcome model $m$.}} Consider $m_{w,x,\cdot}^{(t)} = m_{w,x,\cdot} + t\,\delta m_{w,x,\cdot}$ and let $\partial_t$ denote differentiation w.r.t.\ $t$ at $t=0$. 

\noindent The derivative of $M(\eta)$ w.r.t.\ $t$ as defined in equation \eqref{mfunc} is
\begin{align}\label{expderiv}
& \partial_t M(\eta) \\
& = \sum_{z=1}^L E\Big[ \underbrace{
2 \bar\Theta_{X,z} \,\partial_t \bar\Theta_{X,z}}_I
+ \underbrace{\partial_t \Theta_{X,z}}_{II}
+ \underbrace{2 (\partial_t \bar\Theta_{X,z}) \mathcal{A}^{(Y)}_{X,z} + 2 (\partial_t \bar\Theta_{X,z}) \mathcal{A}^{(D)}_{X,z}}_{III}
+ \underbrace{2 \bar\Theta_{X,z} (\partial_t \mathcal{A}^{(Y)}_{X,z})}_{IV}
 + \underbrace{2 \bar\Theta_{X,z} (\partial_t \mathcal{A}^{(D)}_{X,z})}_{V}
\Big].\notag
\end{align}

First, consider the derivative of the augmentation terms
\begin{align}\label{eq:dA}
\partial_t A^{(Y)}_{X,z}&= - \frac{\delta m_{1,X,z}\,\mathbf{1}\{W=1,Z=z\}}{\pi_{1,z}(X)}  + \frac{\delta m_{1,X,z^-}\,\mathbf{1}\{W=1,Z\neq z\}}{\pi_{1,z^-}(X)},\\
\partial_t A^{(Y)}_{X,z^-}&= - \frac{\delta m_{1,X,z^-}\,\mathbf{1}\{W=1,Z\neq z\}}{\pi_{1,z^-}(X)}
+ \frac{\delta m_{0,X,z^-}\,\mathbf{1}\{W=0,Z\neq z\}}{\pi_{0,z^-}(X)},\notag\\
\partial_t \mathcal{A}^{(Y)}_{X,z}&=\partial_t A^{(Y)}_{X,z} h_{X,z^-}- \partial_t A^{(Y)}_{X,z^-} h_{X,z}.
\end{align}
Taking conditional expectations given \(X\) implies that $E[h_{X,z}|X]=\bar h_{X,z}$ by the augmentation term property $E[A^{(D)}_{X,z}\mid X] = 0$. When additionally considering the  propensity score property $E[\mathbf{1}\{W=w,Z=z\}/\pi_{w,z}(X)|X ]=1$,  we obtain the conditional average derivatives 
\begin{align}
E[\partial_t \mathcal{A}^{(Y)}_{X,z}\mid X] = - \delta \bar g_{X,z}\bar{h}_{X,z^-}+\delta \bar g_{X,z^-}\bar{h}_{X,z}.\label{derivaugment} 
\end{align}
Furthermore, 
\begin{align}
E[\partial_t \mathcal{A}^{(D)}_{X,z}\mid X] = 0.\label{derivaugment2}
\end{align}
Therefore, it follows from the law of iterated expectations and \eqref{derivaugment2} that the expectation of term $V$ in equation \eqref{expderiv} is zero. By the augmentation term property $E[A^{(Y)}_{X,z}\mid X] = E[A^{(D)}_{X,z}\mid X] =0$, the expectation of term $III$ is zero, too

Next, we expand $\Theta_{X,z} = g_{X,z} h_{X,z^-} - g_{X,z^-} h_{X,z}$ into the blocks
\[
\Theta_{X,z} = \bar g_{X,z} \bar h_{X,z^-} + \bar g_{X,z} A^{(D)}_{X,z^-}
+ A^{(Y)}_{X,z} \bar h_{X,z^-} + A^{(Y)}_{X,z} A^{(D)}_{X,z^-}
- \bar g_{X,z^-} \bar h_{X,z} - \bar g_{X,z^-} A^{(D)}_{X,z} 
- A^{(Y)}_{X,z^-} \bar h_{X,z} - A^{(Y)}_{X,z^-} A^{(D)}_{X,z},
\]
and take derivatives w.r.t. $t$:
\begin{align}
\partial_t \Theta_{X,z} = \underbrace{\delta\bar g_{X,z} \bar h_{X,z^-}}_{(a)} + \underbrace{\delta\bar  g_{X,z} A^{(D)}_{X,z^-}}_{(b)} 
+ \underbrace{\partial_t A^{(Y)}_{X,z} \bar h_{X,z^-}}_{(c)} 
+ \underbrace{\partial_t A^{(Y)}_{X,z} A^{(D)}_{X,z^-}}_{(d)}\\
- \underbrace{\delta \bar g_{X,z^-} \bar h_{X,z}}_{(e)} - \underbrace{\delta \bar g_{X,z^-} A^{(D)}_{X,z}}_{(f)} 
- \underbrace{\partial_t A^{(Y)}_{X,z^-} \bar h_{X,z}}_{(g)} - \underbrace{\partial_t A^{(Y)}_{X,z^-} A^{(D)}_{X,z}}_{(h)}.\notag
\end{align}
Taking conditional expectations sets blocks containing an augmentation term to zero, namely $(b,d,f,h)$. Furthermore, making use of \eqref{derivaugment}, $(a)$ and $(c)$ cancel out, as well as $(e)$ and $(g)$. It follows that the expectation of term $II$ in equation \eqref{expderiv} is zero (when also applying the law of iterated expectations).

Finally, we note that 
\begin{align}
\partial_t \bar\Theta_{X,z} = \delta \bar g_{X,z} \bar h_{X,z^-} - \delta \bar g_{X,z^-} \bar h_{X,z},
\end{align}
which enters term $I$ in equation \eqref{expderiv} and corresponds to the negative of equation \eqref{derivaugment}, which enters term $IV$. For this reason, terms $I$ and $IV$ cancel out. 

Therefore, we have that
\begin{align}
\partial_t M(\eta) = 0.
\end{align}

\paragraph{\textit{Perturbations in the treatment model $r$.}} By symmetry (interchanging $Y\leftrightarrow D$), the same arguments demonstrate orthogonality w.r.t.\ perturbations in $r$.

\paragraph{\textit{Perturbations in the propensity score $\pi$.}} Differentiating with respect to $\pi$ affects only the augmentation terms. The derivative introduces terms of the form
\begin{align}
-(Y-m_{w,X,\cdot})\mathbf{1}\{W=w,Z=\cdot\}\frac{\delta\pi_{w,\cdot}(X)}{\pi_{w,\cdot}(X)^2},
\end{align}
whose conditional expectation given $X$ is zero because 
$E[Y-m_{w,X,\cdot}\mid X,W=w,Z=\cdot]=0$. 
Hence the derivative with respect to $\pi$ vanishes in expectation. The same arguments holds for $D$-residuals. Hence any derivative  w.r.t.\ $\pi$ also vanishes.

\paragraph{Step 4: Conclusion.}  

The moment condition
\[
E\Big[ \psi^{\mathrm{CLATE}}(O,\theta_0,\eta \Big]  = 0
\]
is satisfied at the true value of $\eta$, identifying $\theta_0$. Furthermore,
\[
\partial_t E\Big[ \psi^{\mathrm{CLATE}}(O,\theta_0,\eta+t) \Big] \big|_{t=0} = 0,
\]
such that \(\psi^{\mathrm{CLATE}}\) is Neyman-orthogonal at the true value of \(\eta\).

\singlespacing

\section*{Tables} \label{sec:tab}
\addcontentsline{toc}{section}{Tables}

\begin{table}[H]
\centering
\caption{Sample Size by Country and Treatment Status}
\label{tab:sample_country}
\begin{tabular}{lccc}
\toprule
Country & Control & Treated & Total \\
\midrule
ARGE & 266 & 253 & 519 \\
AUSL & 281 & 281 & 562 \\
AUST & 115 & 114 & 229 \\
BELG & 132 & 131 & 263 \\
BRAS & 38 & 40 & 78 \\
CANA & 59 & 58 & 117 \\
CHIL & 29 & 29 & 58 \\
CZEC & 214 & 217 & 431 \\
EIRE & 26 & 26 & 52 \\
FINL & 26 & 27 & 53 \\
GREE & 74 & 76 & 150 \\
HONG & 53 & 55 & 108 \\
HUNG & 52 & 52 & 104 \\
INDI & 102 & 104 & 206 \\
ISRA & 53 & 57 & 110 \\
ITAL & 1557 & 1554 & 3111 \\
NETH & 361 & 350 & 711 \\
NEW  & 224 & 225 & 449 \\
NORW & 263 & 262 & 525 \\
POLA & 377 & 378 & 755 \\
PORT & 190 & 189 & 379 \\
SING & 71 & 68 & 139 \\
SLOK & 43 & 41 & 84 \\
SLOV & 26 & 27 & 53 \\
SOUT & 30 & 32 & 62 \\
SPAI & 232 & 231 & 463 \\
SWED & 313 & 317 & 630 \\
SWIT & 815 & 814 & 1629 \\
TURK & 138 & 140 & 278 \\
UK   & 2882 & 2883 & 5765 \\
USA  & 59 & 57 & 116 \\
\midrule
Total & 9101 & 9088 & 18189 \\
\bottomrule
\end{tabular}
\end{table}

\clearpage

\section*{Figures} \label{sec:fig}
\addcontentsline{toc}{section}{Figures}

\begin{figure}[H]
    \centering
    \caption{Sample size by country (site) and treated vs. control shares.}
    \includegraphics[width=0.8\linewidth]{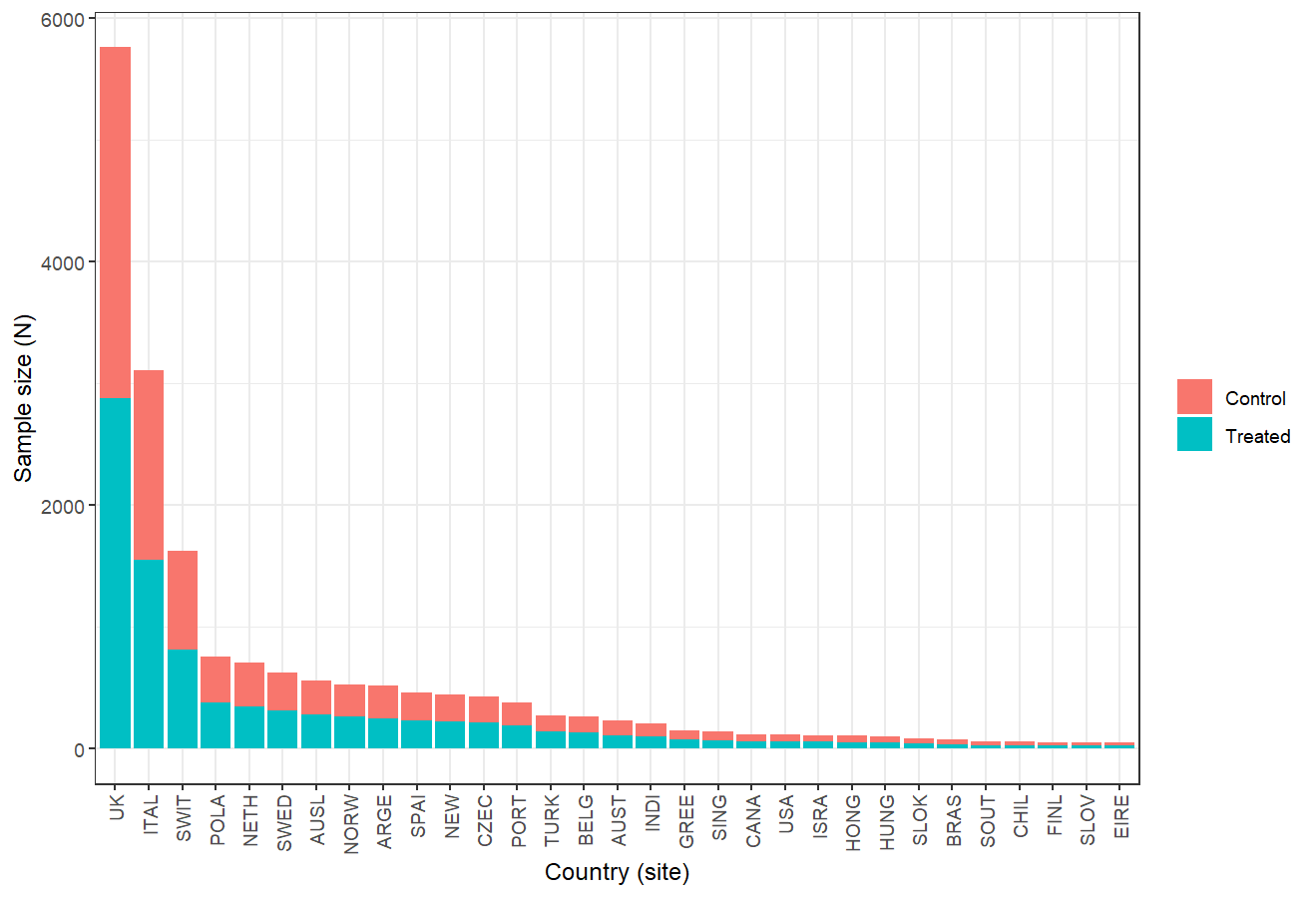}
    \begin{minipage}{0.7\linewidth}
    \footnotesize
    \setstretch{0.95}
    \noindent \textit{Note:} Figure \ref{fig:sites_shares} shows country sample sizes and aspirin assignment shares. Country sizes are uneven, ranging from 52 to 5{,}765 patients (median 229) per country. Nonetheless, treatment assignment is well balanced within countries.
    \end{minipage}
    \label{fig:sites_shares}
\end{figure}

\clearpage

\end{document}

%% file: intro.tex
\section{Introduction} \label{sec:introduction}

Experiments are widely considered the benchmark for credible causal inference because randomization eliminates confounding and delivers unbiased estimates of treatment effects. In practice, however, experimental data often come with important limitations: sample sizes tend to be modest, covariate information is restricted, and external validity is frequently uncertain. By contrast, modern observational data sets are typically much larger and richer, offering detailed high-dimensional covariates and broad population coverage. These features make them attractive for studying treatment effect heterogeneity, but causal interpretation is hindered by the possibility of unobserved confounding. The increasing availability of both experimental and observational data raises an interesting question: to what extent do treatment effect estimates align across data sources once we condition on observed covariates, and what does this reveal about the internal and external validity of identification strategies in experiments and observational studies? 

This paper proposes a framework for testing the homogeneity of conditional average treatment effects (CATEs), given covariates, across multiple experimental and observational studies (sites) such as different regions or countries. When applied solely to randomized experiments, where unobserved confounding is ruled out by design, our test allows assessing treatment effect heterogeneity arising from unobservables that differ across experiments. If CATEs are homogeneous across experiments, this indicates an absence of interactions between the treatment and unobserved factors in the average treatment effect, implying that CATEs are externally valid across experimental settings. Moving beyond experiments, comparing CATEs between experimental and observational data additionally permits assessing the presence of confounding. If the CATE estimates are asymptotically equivalent across experiments and observational studies, then CATEs are unconfounded by unobservables (i.e., internally valid) and homogeneous across settings  (i.e., externally valid). Conversely, if they differ, the discrepancy may stem from unobserved confounding, effect heterogeneity, or both. This may motivate a sequential application of the testing approach; first, across experiments to assess effect homogeneity, and, if homogeneity is not rejected, subsequently across experimental and observational studies to assess unconfoundedness in addition to external validity.

Our testing approach builds on the double machine learning (DML) framework \citep{Chetal2018}, which combines doubly robust (DR) treatment effect estimation \citep{Robins+94, RoRo95, Ha98} with machine learning to flexibly adjust for high-dimensional covariates. More specifically, we extend the \cite{Neyman1959}-orthogonal score function introduced by \cite{apfel2024learningcontrolvariablesinstruments}, who propose a test for whether CATEs within a single study are jointly zero. In contrast, we use an analogous orthogonal formulation to test whether differences in CATEs across studies, experimental and/or observational, are jointly zero. The resulting test is $\sqrt{n}$-consistent (where $n$ denotes the sample size) and asymptotically normal under specific regularity conditions, in particular when the machine learning estimators for the treatment and outcome models converge at rate $o(n^{-1/4})$. 

As an additional methodological contribution, we extend the framework to settings where identification relies on alternative strategies. These include instrumental-variable designs that identify the local average treatment effect (LATE) for compliers whose treatment status responds to the instrument \citep{Imbens+94, Angrist+96}, as well as panel-data settings that rely on parallel trends to identify the average treatment effect on the treated (ATET) based on difference-in-differences \citep{Snow1855}. Because both the conditional LATE and the conditional ATET pertain to specific subpopulations, rather than the full population conditional on covariates, testing homogeneity is informative for evaluating whether these local effects can be extrapolated to the total population, in the spirit of \cite{AnFe2010} and \cite{AronowCarnegie2013}.

We then investigate the finite-sample behavior of our testing approach in a simulation study that mimics settings with multiple experimental and/or observational sites. First, we study the finite-sample size and power of the test in a multi-site experimental design by comparing CATEs across randomized sites under data-generating processes with homogeneous versus heterogeneous CATEs.
Second, we consider a mixed design with randomized treatment in some sites and observational identification in others; we introduce unobserved confounding in the observational sites and evaluate the ability of the test to detect cross-site CATE differences both in the absence and presence of confounding. For each design, we run $1000$ Monte Carlo replications and report summary statistics that describe the performance of our method, such as the rejection rate, standard deviation, and sample size. 

Furthermore, we provide an empirical application for our testing approach using data from the International Stroke Trial, a large multi-country randomized controlled trial which evaluated whether early aspirin allocation to patients with acute ischaemic stroke improved patient health after they had experienced a stoke. We treat countries as separate experimental sites and test whether the conditional effect of randomized aspirin assignment on six-month death or dependency is homogeneous across countries after adjusting for baseline patient characteristics.

A growing literature combines experimental and observational data to improve causal inference \citep{colnet2024causal}. Existing studies use such designs to (i) generalize randomized trial findings to broader populations \citep{cole2010generalizing, Pearl_Bareinboim_2011, ghassami2022combining, stuart2015assessing, pmlr-v180-hatt22a, van2023estimating, park2024informativeness, NBERw33817, imbens2025long, parikh2025doubleML, triantafillou2023learning, lelova2025testing}, (ii) increase statistical efficiency, especially for heterogeneous treatment effects \citep{yang2020improved, wu2022integrative, yang2023elastic, cheng2021adaptive, yang2025cross, rosenman2022propensity, rosenman2023combining, hatt2022combining, Brantner_2024_multipleRCT, epanomeritakis2025choosing}, and (iii) diagnose or correct bias in observational analyses by using experimental benchmarks \citep{kallus2018removing, yang2020improved, wu2022integrative, yang2023elastic, liu2025causal, parikh2025doubleML, triantafillou2023learning, lelova2025testing, chen2025causal}. 

We complement these efforts by introducing a framework that tests whether CATEs are homogeneous across multiple experimental and observational sites. By comparing CATEs across experiments, the test can detect heterogeneity across experimental sites driven by unobservables and evaluate the external validity of experimental CATEs. While by comparing CATEs from experiments and observational sites, our test can diagnose the presence of hidden confounding and assess internal validity of treatment effects. Therefore, our approach provides a unified way to assess both internal and external validity of CATEs using distinct research designs.

The remainder of this paper is organized as follows. Section \ref{sec:literature} provides a detailed literature survey on studies combining multiple experimental and observational data for treatment effect evaluation. Section \ref{sec:method} introduces the identifying assumptions and outlines our method. Section \ref{sec:method2} extends the testing framework to instrumental and panel data contexts. Section \ref{sec:simulation} provides a simulation study that investigates the finite sample performance of our proposed test. Section \ref{sec:data} illustrates the method using the International Stroke Trial. Section \ref{sec:discussion} concludes. 

%% file: literature.tex
\section{Literature survey} \label{sec:literature}

Our study contributes to the three strands of research mentioned above by combining experimental and observational evidence to address unobserved confounding and to evaluate the internal as well as the external validity of CATEs. Within this body of work, the literature largely follows two approaches: (i) one explicitly models and estimates a confounding function by pooling information from RCTs and observational datasets, (ii) the other develops methods that combine or pool CATE estimators from experimental and observational samples using adaptive weights chosen to balance bias and efficiency.

The first approach focuses on the confounding function, defined as the conditional gap between causal and observational treatment effects, as a central object of interest. Early contributions such as \cite{kallus2018removing} propose a method that first learns an observational CATE function and then estimates a low-dimensional correction term using experimental data, so that the observational estimate matches the randomized benchmark even with only partial covariate overlap. Building on this idea, \citet{yang2020improved} introduce a data fusion framework in which both the CATE and the confounding function are identifiable once observational and experimental samples are coupled. They show that their method improves efficiency relative to cases where there are only experimental samples. Subsequent work by \citet{wu2022integrative} extend this framework by proposing an R-learner that incorporates flexible machine learning methods to approximate the CATE, the confounding function, and other nuisance components. 

Even more recently, \cite{yang2023elastic} introduce a test-based elastic approach for integrating trial and real-world data. Their method first uses the RCT as a benchmark to test whether the observational sample suffers from bias. If the test fails, only the data from the RCT is used, but if the test supports comparability, the two sources are combined for efficiency. Complementing these elastic integration ideas, \citet{liu2025causal} develop a direct hypothesis test for unconfoundedness by comparing treatment–outcome contrasts estimated from the RCT and the observational data. Unlike \citet{yang2023elastic}, which couples a pretest with an adaptive estimator, \citet{liu2025causal} focuses on diagnosis by flagging when the observational sample is likely confounded before applying fusion or machine learning estimators that assume ignorability. \citet{parikh2025doubleML} push this diagnostic idea further by asking which assumption breaks when the two sources disagree: they develop a double machine learning framework by introducing a statistical quantity that distinguishes failures of ignorability in the observational sample from failures of external validity of the experiment. 

A complementary Bayesian approach incorporates uncertainty about when observational data are safe to use. \cite{triantafillou2023learning} propose Bayesian CATE estimation that adaptively borrows from observational data, while \cite{lelova2025testing} study identification and transportability of CATEs under an unknown causal graph when combining experimental and observational samples. Beyond the econometric literature, related work in computer science by \cite{hatt2022combining} proposes a representation learning framework which first learns the shared covariate structure from observational data and then uses experimental data to calibrate the estimation of treatment effects. They formalize the bias from unmeasured confounding as a confounding function, learn this bias by comparing observational and experimental predictions, and then use it to debias CATE estimates. 

The second strand of research avoids modeling a confounding function and instead combines CATE estimators computed separately in experimental and observational samples. \citet{cheng2021adaptive} combine kernel-based CATE estimates using data-driven weights that default to the experimental data when bias is suspected and combine both experimental and observational sources when estimates align. \citet{yang2025cross} generalize this idea by choosing the weights given to each data source through cross-validation in a joint loss framework, trading off bias and variance across the two sources.

Related work by \cite{rosenman2022propensity} propose to combine RCTs and observational data by stratifying on the observational propensity score and placing experimental units into the same strata. Within each stratum, they estimate treatment effects from experimental and observational sources and then merge them either by spiking experimental data into observational bins or through a data-driven weighting scheme that balances bias and variance. In subsequent work, \cite{rosenman2023combining} extends this approach using Stein-type shrinkage estimators. Their method adaptively shrinks observational estimates toward unbiased experimental estimates. They show that these estimators reduce the mean squared error relative to performing only experimental analyzes. While this line of work focuses on optimally combining experimental and observational evidence to improve efficiency, it implicitly assumes that treatment effects are sufficiently stable across settings.

In contrast, \citet{chen2025causal} investigates whether causal machine learning methods can produce reliable CATE estimates using data from two large RCTs. They show that individualized treatment effects derived from a wide range of machine learning methods fail to replicate across training and test splits or across trials, even in the absence of confounding. This highlights the difficulty of obtaining externally valid CATE estimates and the need for systematic approaches to test the stability of CATEs across settings. 

For a broader synthesis of the literature, \citet{colnet2024causal} provides a systematic review of approaches that integrate experimental and observational data. Additionally, \citet{Brantner2023Integration} provide a review of methods to combine multiple RCTs or RCTs with observational data, with a focus on treatment effect heterogeneity. They classify approaches by the type of data available and discuss both parametric and machine learning strategies for estimating CATEs. A key takeaway is that comparing CATEs across sources provides a way to assess stability and detect potential confounding. 

Despite advances in combining experimental and observational data, most applications focus on a single RCT paired with one observational dataset. As a result, little is known about whether CATEs align across multiple experiments and observational sources. An exception is \citet{Brantner_2024_multipleRCT}, develop and who study methods for estimating CATEs when several RCTs are available. They adapt S-learner, X-learner, and causal forest estimators to the multi-trial setting. They show that strategies allowing trial-level heterogeneity outperform naive pooling that ignores study differences. Their work highlights the challenges of integrating multiple experiments, but does not examine the alignment of CATEs between experimental and observational data. Our study fills this gap by focusing on testing the homogeneity of CATEs across both experimental and observational sources.

%% file: method.tex
\section{Assumptions and testing approach} \label{sec:method}

$D$ denotes the binary treatment and $Y$ the outcome of interest. Using the potential outcomes framework as advocated in \cite{Neyman23} and \cite{Rubin74}, we denote by $Y(d)$ the potential outcome when exogenously setting the treatment $D$ of a subject to value $d \in {1,0}$. More generally, we will use capital letters for random variables and lower case letters for their realizations. By representing the potential outcome $Y(d)$ as a function solely of a subject’s own treatment status $D=d$, we implicitly assume that the potential outcomes of one subject are not influenced by the treatment status of others. This assumption is known as the stable unit treatment value assumption (SUTVA), see \cite{Rubin80} and \cite{Cox58}, and is invoked throughout. Furthermore, let $X$ denote a set of observed pretreatment covariates, and let $Z$ be a discrete variable that indexes different setups or studies in which the experimental or observational data were collected (for example, sites or regions). The variable can take integer values $z \in {1,...,L}$, with $L$ denoting the number of setups.

We suggest a method to test effect homogeneity in conditional average treatment effects (CATEs) across different experiments, or selection-on-observables (and effect homogeneity) across  experimental and observational data, respectively. First, we consider the case of comparisons within experimental studies. Suppose that treatment is randomly assigned within each site or region $Z$, possibly conditional on covariates $X$ (as in stratified randomization). This corresponds to the standard selection-on-observables assumption, also known as unconfoundedness or conditional independence \citep{Im04}.
\begin{assumption}[Conditional independence of the treatment]\label{ass1}
\begin{eqnarray*}
\{Y(1),Y(0)\} {\perp\!\!\!\perp} D | X, Z, 
\end{eqnarray*}
\end{assumption}
where ${\perp\!\!\!\perp}$ denotes statistical independence. 

In addition, we require a condition ensuring that treated and untreated units are observed in all relevant subpopulations of $X$ and $Z$:
\begin{assumption}[Common support]\label{ass2}
\begin{eqnarray*}
0<\Pr(D=d,Z=z|X)<1,\textrm{ }\forall d \in \{1,0\}\textrm{ and } z \in \{1,...,L\}.
\end{eqnarray*}
\end{assumption}
The common support assumption guarantees overlap in the treatment assignment across different experiments and covariate profiles. It rules out situations where, conditional on covariates $X$, treatment assignment or assignment to a specific experiment is deterministic. 

The conditional average treatment effect (CATE) given covariates $X$ and experiment $Z$ is defined as 
\begin{align}
\Delta_{x,z}=E[Y(1)-Y(0)|X=x,Z=z]
\end{align}

Under (conditional) treatment randomization, which implies the satisfaction of Assumption \ref{ass1}, we have that $\Delta_{X,Z}$ corresponds to 
\begin{align}
\delta_{x,z}=E[Y|D=1,X=x,Z=z]-E[Y|D=0,X=x,Z=z].
\end{align}

We are interested in testing whether these conditional effects vary across experiments. Formally, we impose:
\begin{assumption}[Conditional effect homogeneity]\label{ass3}
\begin{eqnarray*}
E[Y(1)-Y(0)|X,Z]=E[Y(1)-Y(0)|X].
\end{eqnarray*}
\end{assumption}
This assumption states that CATEs are homogeneous across experiments $Z$. Such homogeneity may hold for two distinct reasons. First, treatment effects may not interact with unobserved heterogeneity once we condition on $X$. An example where this is satisfied is the following structural model:
\begin{align}\label{addsep}
Y = \kappa(D,X) + \eta(U),
\end{align}
where $U$ denotes unobserved characteristics and $\kappa$ and $\eta$ are unknown functions. While the effect of $D$ may vary arbitrarily across $X$, it does not vary with $U$ conditional on $X$ due to the additive separability of $\kappa$ and $\eta$. In such cases, even if unobserved characteristics differ across experiments, they do not generate treatment effect heterogeneity.

Second, treatment effects may indeed interact with unobserved heterogeneity, but the distribution of this heterogeneity remains stable across experiments. For instance, consider the structural model
\begin{align}\label{noaddsep}
Y = \kappa(D,X,U),
\end{align}
where the effect of $D$ on $Y$ may arbitrarily interact with $U$. This second case, effect heterogeneity existing but not being detectable because the distribution of $U$ is identical across experiments, appears less plausible when experimental sites differ substantially in institutional, geographic, or temporal contexts. Such differences typically shift the distribution of unobserved characteristics. The more variation there is in unobserved heterogeneity across experiments, the more informative the data become for detecting interactions between treatment effects and unobservables.

Conditional on Assumption \ref{ass1}, Assumption \ref{ass3} yields the following testable null hypothesis:
\begin{align}\label{H0}
H_0: \underbrace{\mu_{1,x,z}-\mu_{0,x,z}}_{\delta_{x,z}} - (\underbrace{\mu_{1,x,z'}-\mu_{0,x,z'}}_{\delta_{x,z'}})=0, \quad\forall z,z' \in \{1,..,L\}\textrm{ and }x \in \mathcal{X},
\end{align}
where $\mathcal{X}$ denotes the support of $X$ and $\mu_{d,x,z}=E[Y|D=d,X=x,Z=z]$ denotes the conditional mean outcome.

To test the null hypothesis in \eqref{H0}, we adapt the doubly robust conditional independence test of \citet{apfel2024learningcontrolvariablesinstruments} to the problem of effect homogeneity. That is, we extend their approach, which tests whether CATEs differ from zero, to instead test whether differences in CATEs across experiments are equal to zero. Note that \eqref{H0} can equivalently be written as
\begin{align}\label{hypothesis2}
  H_0: \theta_0= E\left[\sum_{z=1}^L[(\mu_{1,x,Z=z}-\mu_{0,x,Z= z}-\mu_{1,x,z^-}+\mu_{0,x,z^-})^2+(\mu_{1,x,Z=z}-\mu_{0,x,Z= z}-\mu_{1,x,z^-}+\mu_{0,x,z^-})]\right]=0,
\end{align}
where $z^-$ denotes values of $Z$ different from $z$, such that $Z \neq z$. 

Denote by $p_{d,z}(X)=\Pr(D=d,Z=z|X)$ the joint propensity of treatment and being in a specific experiment or site. We denote the nuisance parameters by 

$\eta=(p_{1,z}(X),p_{1,z^-}(X),\mu_{1,X,z},\mu_{1,X,z^-},p_{0,z}(X),p_{0,z^-}(X),\mu_{0,X,z},\mu_{0,X,z^-})$. 

Modifying \cite{apfel2024learningcontrolvariablesinstruments}, who consider simple difference in conditional means to test whether any CATE is different from zero, to double (or differences in) differences, to test whether any difference in CATEs (across experiments) is different from zero. More concisely, testing with a multivalued $Z$ can be based on the following score function, in which $O=(Y,D,X,Z)$ denotes the random variables:
\begin{align}\label{score1}
&\quad \psi(O,\theta,\eta)\\
&=\sum_{z=1}^L(\mu_{1,X,z}-\mu_{0,X,z}-\mu_{1,X,z^-}+\mu_{0,X,z^-})^2\notag\\&
+\sum_{z=1}^L2(\mu_{1,X,z}-\mu_{0,X,z}-\mu_{1,X,z^-}+\mu_{0,X,z^-})\notag\\
&\left(\frac{(Y-\mu_{1,X,z}) 1(D=1,Z=z)}{p_{1,z}(X)}-\frac{(Y-\mu_{0,X,z}) 1(D=0,Z=z)}{p_{0,z}(X)}\right.\notag\\
&-\left.\frac{(Y-\mu_{1,X,z^-})1(D=1,Z\neq z)}{p_{1,z^-}(X)}+\frac{(Y-\mu_{0,X,z^-})1(D=0,Z\neq z)}{p_{0,z^-}(X)}\right)\notag\\&
+\sum_{l=1}^L(\mu_{1,X,z}-\mu_{0,X,z}-\mu_{1,X,z^-}+\mu_{0,X,z^-})\notag\\&
+\sum_{z=1}^L\left(\frac{(Y-\mu_{1,X,z}) 1(D=1,Z=z)}{p_{1,z}(X)}-\frac{(Y-\mu_{0,X,z}) 1(D=0,Z=z)}{p_{0,z}(X)}\right.\notag\\
&-\left.\frac{(Y-\mu_{1,X,z^-})1(D=1,Z\neq z)}{p_{1,z^-}(X)}+\frac{(Y-\mu_{0,X,z^-})1(D=0,Z\neq z)}{p_{0,z^-}(X)}\right)-\theta\notag.
\end{align}
This score has a variance that is bounded away from zero, is zero in expectation under the null hypothesis in equation \eqref{hypothesis2} when $\theta_0=0$, and is Neyman-orthogonal, see proof provided in Appendix of \cite{apfel2024learningcontrolvariablesinstruments}. This follows directly from the proofs in \cite{apfel2024learningcontrolvariablesinstruments}, as our score function is based on applying their type of score function twice to turn it into a double (rather than a single) difference across $\mu_{d,X,z}$. As a the double difference is just a linear combination of the single differences, the asymptotic findings in \cite{apfel2024learningcontrolvariablesinstruments} directly apply to our case, too. 
In particular, cross-fitted estimators of $\theta_0$ based on the score function \eqref{score1} is asymptotically normal and $\sqrt{n}-$ consistent under specific regularity conditions, in particular if machine learners used for estimating nuisance parameters $\eta$ have a convergence rate of $o(n^{-1/4})$. 

The same testing framework can also be applied for comparing experimental and observational studies in a second step following the within-experiments comparison. In purely observational data, where Assumption \ref{ass1} cannot be taken for granted, rejection of \eqref{H0} or \eqref{hypothesis2} may reflect violations of Assumption \ref{ass1}, Assumption \ref{ass3}, or both. However, if experimental data suggest that CATEs are homogeneous, i.e., Assumption \ref{ass3} holds, then comparisons of CATEs between experimental and observational studies provide a direct test of Assumption \ref{ass1}. Specifically, one can define $Z$ such that $Z=1$ indicates observations from experimental studies, while $Z=2,\dots,L$ index observational studies. Alternatively, $Z$ may distinguish between observational studies only. In both cases, testing can again be implemented using the score in \eqref{score1}.

\section{Alternative identifying assumptions} \label{sec:method2}

Our method for testing effect homogeneity can also be adapted to settings where treatment is not conditionally exogenous. For instance, access to treatment may be randomly assigned conditional on $X$, while actual treatment take-up deviates from assignment due to noncompliance. In this case, we may use assignment, henceforth denoted by $W$, as an instrument for actual treatment $D$. Following \cite{Imbens+94} and \cite{Angrist+96}, we denote by $D(w)$ the potential treatment as a function of instrument $W$ and by $Y(w,d)$ the potential outcome as a function of $W$ and $D$. We impose the following instrumental variable (IV) assumptions conditional on covariates $X$
and experiment $Z$, see \cite*{Abadie00}:
\begin{assumption}[IV assumptions]\label{ass4}
\begin{align*}
&\{D(w), Y(w',d)\} {\perp\!\!\!\perp} W |X,Z\textrm{ for }w,w',d\in\{0,1\},\quad \Pr(Y(1,d)=Y(0,d)=Y(d)|X,Z)=1,\notag\\
&\Pr(D(1)\ge D(0)|X,Z)=1,\quad  E[D|W=1,X,Z]-E[D|W=0,X,Z]\neq 0,\\
&0<\Pr(W=1|X,Z)<1.\notag
\end{align*}
\end{assumption}
The first line of Assumption \ref{ass4} requires that the instrument is as good as randomly assigned and satisfies the exclusion restriction conditional on $X$ and $Z$. The second line rules out the existence of defiers, but it also requires the existence of compliers conditional on $X$, due to the nonzero conditional first stage. The third line imposes common support on the instrument, implying that assignment is not deterministic in $X$ and $Z$. 

Assumption \ref{ass4} permits identifying conditional local average treatment (CLATE) effect among the subgroup compliers, denoted by $c$, who are treated only if the instrument is equal to one: $c: D(1)=1,D(0)=0$. The CLATE given covariates $X$ and experiment $Z$ is defined as  
\begin{align}
\Delta_{c,x,z}=E[Y(1)-Y(0)|D(1)=1,D(0)=0,X=x,Z=z].
\end{align}
The CLATE is identified using a Wald-type estimand \citep{Wald40}, defined as the ratio of the reduced-form effect of the instrument on the outcome to the first-stage effect of the instrument on the treatment, conditional on $X$ and $Z$:
\begin{align}
\delta_{c,x,z}=\frac{E[Y|W=1,X=x,Z=z]-E[Y|W=0,X=x,Z=z]}{E[D|W=1,X=x,Z=z]-E[D|W=0,X=x,Z=z]}= \frac{g_{x,z}}{h_{x,z}},
\end{align}
where we define the reduced-form and first-stage effects as $\bar g_{x,z} = m_{1,x,z} - m_{0,x,z}$ and $\bar h_{x,z} = r_{1,x,z} - r_{0,x,z}$, 
with $m_{w,x,z} = E[Y\mid W=w,X=x,Z=z]$ and $
r_{w,x,z} = E[D\mid W=w,X=x,Z=z]$.

Considering, in analogy to equation \eqref{H0} for the CATE, the following null hypothesis,
\begin{align}\label{H0CLATE}
H_0: \delta_{c,x,z}-\delta_{c,x,z'}=0, \quad\forall z,z' \in \{1,..,L\}\textrm{ and }x \in \mathcal{X},
\end{align}
permits testing the following effect homogeneity assumption among compliers: 
\begin{assumption}[Conditional effect homogeneity among compliers]\label{ass5}
\begin{align*}
E[Y(1)-Y(0)|D(1)=1,D(0)=0,X,Z]=E[Y(1)-Y(0)|D(1)=1,D(0)=0,X].
\end{align*}
\end{assumption}

Since $\delta_{c,x,z} = g_{x,z}/h_{x,z}$, we note that the null hypothesis \eqref{H0CLATE} can be equivalently written in cross-multiplied form as
\begin{align}
\bar \Theta_{x,z} =\bar g_{x,z}  \bar h_{x,z'} -  \bar g_{x,z'} \bar h_{x,z} = 0, \quad\forall z,z' \in \{1,..,L\}\textrm{ and }x \in \mathcal{X}.
\end{align}
Hence, a ratio-free version of the null hypothesis for the CLATE that is analogous to equation \eqref{hypothesis2} for the CATE is given by
\begin{align}\label{hypothesisCLATE}
H_0: \theta_0 = E\!\left[\sum_{z=1}^L \Big((\bar g_{X,z} \bar h_{X,z^-}-\bar g_{X,z^-}\bar h_{X,z})^2 + (\bar g_{X,z}\bar h_{X,z^-}-\bar g_{X,z^-}\bar h_{X,z})\Big)\right] = 0,
\end{align}
where $z^-$ denotes values of $Z$ different from $z$. We also define the propensity score $\pi_{w,z}(x) = \Pr(W=w,Z=z\mid X=x)$ and collect the nuisance parameters in $\eta = (m_{w,x,z},r_{w,x,z},\pi_{w,z}(x),\pi_{w,z^-}(x))_{w\in\{0,1\},z\in\{1,\dots,L\}}$. 

Following a similar logic as in the CATE case, we construct a DR score function based on \eqref{hypothesisCLATE} for the CLATE setting. To this end, define the DR augmentations for the reduced form and first stage effects as
\begin{align*}
g_{X,z} &= 
m_{1,X,z}-m_{0,X,z}
+\frac{(Y-m_{1,X,z})1(W=1,Z=z)}{\pi_{1,z}(X)}
-\frac{(Y-m_{0,X,z})1(W=0,Z=z)}{\pi_{0,z}(X)}\\
h_{X,z} &= 
r_{1,X,z}-r_{0,X,z}
+\frac{(D-r_{1,X,z})1(W=1,Z=z)}{\pi_{1,z}(X)}
-\frac{(D-r_{0,X,z})1(W=0,Z=z)}{\pi_{0,z}(X)}\\
\end{align*}
Using these, we denote the cross-product difference by
\begin{align*}
\Theta_{X,z} = g_{X,z}h_{X,z^-} - g_{X,z^-}h_{X,z}.
\end{align*}
Denoting by $O=(Y,D,X,Z,W)$ the random variables, a DR score function for testing the null hypothesis in \eqref{hypothesisCLATE} is
\begin{align}\label{scoreCLATE}
\psi^{\text{CLATE}}(O,\theta,\eta)
&= \sum_{z=1}^L \left(\bar\Theta_{X,z}^2 + \Theta_{X,z}\right)\\
&\quad + \sum_{z=1}^L 2\,\bar\Theta_{X,z}
\left[h_{X,z^-}\left(\frac{(Y-m_{1,X,z})1(W=1,Z=z)}{\pi_{1,z}(X)}
-\frac{(Y-m_{0,X,z})1(W=0,Z=z)}{\pi_{0,z}(X)}\right)  \right.\notag\\
&\quad -h_{X,z}\left(\left.\frac{(Y-m_{1,X,z^-})1(W=1,Z\neq z)}{\pi_{1,z^-}(X)}
+\frac{(Y-m_{0,X,z^-})1(W=0,Z\neq z)}{\pi_{0,z^-}(X)}\right)\right]\notag\\
&\quad + \sum_{z=1}^L 2\,\bar\Theta_{X,z}
\left[g_{X,z^-}\left(\frac{(D-r_{1,X,z})1(W=1,Z=z)}{\pi_{1,z}(X)}
-\frac{(D-r_{0,X,z})1(W=0,Z=z)}{\pi_{0,z}(X)} \right) \right.\notag\\
&\quad -g_{X,z}\left(\left.\frac{(D-r_{1,X,z^-})1(W=1,Z\neq z)}{\pi_{1,z^-}(X)}  +\frac{(D-r_{0,X,z^-})1(W=0,Z\neq z)}{\pi_{0,z^-}(X)}\right)\right]
-\theta.\notag
\end{align}
This score function has zero mean under the null hypothesis $H_0:\theta_0=0$ and is Neyman-orthogonal with respect to the nuisance parameters $\eta$, see the proof provided in Appendix \ref{sec:appendixa}. Consequently, cross-fitted estimators of $\theta_0$ based on \eqref{scoreCLATE} are $\sqrt{n}$-consistent and asymptotically normal under specific regularity conditions, in particular if machine learning estimators of the nuisance functions converge at rate $o(n^{-1/4})$.

We note that, although the construction of the CLATE score function is related to the approach of \citet{apfel2024learningcontrolvariablesinstruments}, there is a conceptual difference compared to the CATE framework considered in their paper and in our Section \ref{sec:method}. While the DR function for the CATE is linear (but not quadratic) in the debiasing terms, in which outcome regression residuals are reweighted by the inverse of propensity scores (also known as augmented residuals), the CLATE score involves a cross-product between the DR estimators of the reduced form and first-stage effects. Consequently, the CLATE moment condition contains both $\Theta_{X,z}$ and its square, reflecting the bilinear structure of the CLATE, which depends on the ratio of two conditional effects. This introduces second-order terms in the score but does not alter Neyman orthogonality, as the influence of the nuisance parameters still cancels out through the residual orthogonality conditions $E[Y - m_{w,X,z}\mid X,Z,W] = 0$ and $E[D - r_{w,X,z}\mid X,Z,W] = 0$. In this sense, the CLATE score extends the construction of scores based on squared differences in regression functions to a setting where both the numerator (reduced form) and denominator (first stage) of the parameter of interest must be debiased simultaneously.

As for Assumption \ref{ass3} in the CATE case, it is worth noting that Assumption \ref{ass5} may hold for two reasons: either CLATEs do not depend on unobservables, or the distribution of unobservables is stable across experiments. To illustrate, consider the outcome model in equation \eqref{noaddsep} together with a threshold-crossing treatment model
\begin{align}\label{treatmodel}
D=I\{\lambda(Z,X)>\eta(V)\},
\end{align}
where $I\{\ \cdot\}$ is the indicator function that is equal to one if its argument is satisfied and zero otherwise, $\lambda$ and $\eta$ are unknown functions, and $V$ are unobservables affecting the treatment. 

We note that the threshold-crossing model for treatment assignment in equation \eqref{treatmodel} both implies and is implied by treatment monotonicity, as shown in \citet{Vy02}. Regarding effect heterogeneity, the unobservable $V$ may be arbitrarily associated with $U$ under our IV assumptions, so that heterogeneity of treatment effects in $U$ generally also induces heterogeneity with respect to $V$ and hence across compliance types, defined by whether $I\{\lambda(1,X)>\eta(V)\}=1$ and $I\{\lambda(0,X)>\eta(V)\}=0$. Therefore, if it can be assumed that unobservables differ across $Z$, then satisfaction of Assumption \ref{ass5} points to homogeneous effects. This, in turn, implies that treatment effects do not depend on compliance behavior - which is itself a function of unobservables - conditional on $X$, an assumption discussed in \citet{AnFe2010} and \citet{AronowCarnegie2013}:
\begin{assumption}[CLATE equals CATE]\label{ass:cond_effect_homogeneity}
\begin{align*}
	E[Y(1)-Y(0)|D(1),D(0),X,Z]=E[Y(1)-Y(0)|X].
\end{align*}
\end{assumption}
An important implication of this assumption is that it allows extrapolating the CLATE to the entire population, since under Assumption \ref{ass:cond_effect_homogeneity} the CLATE coincides with the CATE. In other words, the identified effect is no longer local to compliers, but represents the average conditional effect for the full population.

Further, alternative identifying assumptions can be considered when panel data (or also repeated cross sections) are available, in which outcomes are observed both before and after the introduction of treatment. To this end, we introduce time index $t \in {0,1}$, where $t=0$ refers to the pre-treatment period and $t=1$ to the post-treatment period, to denote by $Y_t$ and $Y_t(d)$ the outcome and the potential outcome (given $D=d$) at time $t$, respectively. This setting permits effect identification based on the parallel trends assumption, which requires conditional independence in outcome trends rather than in outcome levels (as imposed by Assumption \ref{ass1}). A set of sufficient assumptions for identifying the conditional average treatment effect on the treated (CATET) in panel data based on the difference-in-differences (DiD) approach is the following, see, e.g., \cite*{Abadie2005}; \cite*{Lechner2010}:
\begin{assumption}[DiD assumptions]\label{ass6}
\begin{align*}
&E[Y_1(0)-Y_0(0)|D=1,X,Z]=E[Y_1(0)-Y_0(0)|D=0,X,Z], \\
&E[Y_0(1)-Y_0(0)|D=1,X,Z]=0, \notag \\
&\Pr(D=1|X, Z)<1. \notag
\end{align*}
\end{assumption}
The first condition in Assumption \ref{ass6} formalizes the conditional common trends assumption: given (presumably exogenous) covariates $X$ and experiment $Z$, no unobserved factors simultaneously affect both treatment assignment and the trend of mean potential outcomes under non-treatment. In DiD settings, it is worth noting that in the context of DiD, multiple experiments $Z$ may for instance correspond to multiple treated regions observed within the same dataset. The second condition rules out average anticipation effects among the treated, conditional on $X$. It requires that treatment status $D$ does not causally influence pretreatment outcomes in expectation of the treatment to come. The third line imposes a specific common support condition for identifying the CATET, requiring that for every covariate profile $X$ and experiment $Z$ observed among the treated, there also exist some untreated observations with the same $(X,Z)$. 

When replacing $Y$ by the outcome difference $Y_1-Y_0$ in the definitions of the conditional mean outcomes $\mu_{D,X,Z}$ introduced in Section \ref{sec:method}, the null hypotheses \eqref{H0} and \eqref{hypothesis2}, as well as the score function \eqref{score1}, can be redefined to evaluate effect heterogeneity across CATETs in different experiments. A natural question is whether these effects can be extrapolated to the total population, i.e., whether the CATET coincides with the CATE. In general, this is not the case because the parallel trends condition in Assumption \ref{ass6} is only imposed for the untreated potential outcomes, such that identification is restricted to the treated group. In particular, treatment effects may differ with respect to time-invariant confounders that are allowed to differ between treated and untreated units. However, if experiments differ in such time-invariant confounders, and effect homogeneity across experiments is not rejected, this suggests that treatment effects do not depend on them. In this case, the CATET coincides with CATE, as expressed in the following assumption:
\begin{assumption}[CATET equals CATE]\label{ass:cond_effect_homogeneity2}
\begin{align*}
	E[Y(1)-Y(0)|D=1,X,Z]=E[Y(1)-Y(0)|X].
\end{align*}
\end{assumption}

The following structural model illustrates the role of time-invariant unobservables. Let
\begin{align}
Y_t= \kappa_t(D,X) + \eta(D,U) + \varepsilon_t,
\end{align}
where $\kappa_t(D,X)$ is an unknown, time-varying function of covariates $X$ and treatment $D$, $\eta(D,U)$ is a time-invariant function of unobservables $U$ that may interact with treatment, and $\varepsilon_t$ is an idiosyncratic, time-varying error. For the potential outcome under non-treatment,
\begin{align}
Y_t(0)= \kappa_t(0,X) + \eta(0,U) + \varepsilon_t,
\end{align}
it follows that differencing across time, $Y_1(0)-Y_0(0)$, eliminates $\eta(0,U)$ due to its additive separability. Hence, the parallel trends condition holds with respect to $Y_t(0)$ conditional on $X$, even if the distribution of $U$ differs between treatment groups. However, treatment effects may still vary across groups, since arbitrary interactions between $D$ and $U$ are allowed in $\eta(D,U)$. Now consider the alternative model
\begin{align}
Y_t= \kappa_t(D,X) + \eta(U) + \varepsilon_t,
\end{align}
which rules out such interactions and implies additive separability of $U$ and $D$. In this case, treatment effects are homogeneous in $U$, as in classical linear panel regression models. Therefore, if $U$ plausibly varies across experiments $Z$ but CATETs are found to be constant across $Z$ (and thus across distributions of $U$), this provides evidence in favor of Assumption \ref{ass:cond_effect_homogeneity2}, which justifies extrapolating effects identified for the treated (CATET) to the entire population (CATE).

%% file: simulation.tex
\section{Simulation Study} \label{sec:simulation}

This section describes a simulation study to investigate the finite sample behavior of our proposed test of homogeneity of CATEs across experimental and observational sites. We first consider comparisons across experiments and and base our simulations on the following data generating process (DGP): 

\begin{align*}
    Y &= D + DX'\beta + \delta DZ + X'\beta + U, \\
    D &\sim \text{Bernoulli}(q), \\
    X &\sim \mathcal{N}(0, \Sigma), \\
    Z& \sim \text{Bernoulli}(\pi), \\
    U &\sim \mathcal{N}(0,1)
\end{align*}

Where the outcome $Y$ is a function of the treatment $D$, $X$ the covariates, $Z$ is an indicator of an experimental site and $U$ denotes the error term. In the experimental design, $D$ is randomly assigned based on a Bernoulli distribution with probability $q$ and Assumption~\ref{ass1} (conditional independence) holds by construction. $X$ is a vector of covariates of dimension $p$, drawn from a multivariate normal distribution with zero mean and covariance matrix $\Sigma$. In this specification, $\Sigma$ equals the identity matrix, implying that all covariates are independent and have unit variance. $Z$ is an indicator of an experimental (or observational) site, generated independently from a Bernoulli distribution with probability $\pi$. The coefficients $\beta$ determine the impact of the covariates $X$ on $Y$. Finally, $U$ is a random and normally distributed error term. 

In the observational design, the element which changes is the  treatment assignment which is no longer randomized:

\begin{align*}
    D_{obs}& = \mathbb{I}\{ X'\beta + \rho U + V > 0 \}, \\
    V &\sim \mathcal{N}(0,1)
\end{align*} 

Thus, we consider the case where treatment $(D_{obs})$ depends on covariates $X$ and the error terms $U$ and $V$. Where $\rho$ determines the strength of confounding in the observational sites. At the same time, the parameter $\delta$, from the outcome equation, governs the degree of effect heterogeneity across experimental and/or observational sites. Consequently, when $\delta=0$, treatment effects are constant across sites, while $\delta \neq 0$ induces heterogeneity in CATEs across sites, indexed by $Z$. Likewise, when $\rho=0$, there is no confounding in the observational sites, while $\rho \neq 0$ introduces confounding. 

We implement a cross-fitted, doubly robust (DR) score test based on double difference score in \eqref{score1}, adapted from \cite{apfel2024learningcontrolvariablesinstruments} to test whether CATEs are homogeneous across sites $Z$. For each $Z$, we estimate conditional means and joint propensities by a penalized lasso regression using five fold $k=5$ cross fitting with default parameters as in the \texttt{glmnet} package in R. We then build DR residuals and combine them into a double difference score across individual sites $z$. To ensure overlap, we trim the estimated conditional probabilities below $0.05$ and above $0.95$.

The overall test statistic is computed as the sample mean of the individual scores over the retained sample after trimming. In addition, the standard error is obtained from the score variance scaled by the sample size. Finally, a normal approximation is used to compute $p$-values. The simulation scenarios vary in several dimensions. We consider three sample sizes, $n = 500, 2000$ and $8000$. In our main specification, we set \( k = 5 \), \( l = 2 \), \( p = 100 \), use Lasso as the machine learner, and fix the trimming threshold at \( \varepsilon = 0.05 \). We perform $R=1000$ Monte Carlo replications. To assess size, we impose the null of cross-site homogeneity by setting $\delta=0$. To assess power, we introduce cross-site heterogeneity by setting $\delta=1$. In the mixed design, we further allow for unobserved confounding in observational sites by setting $\rho\in\{0,0.5\}$, so that we study the performance of the test under both unconfounded and confounded observational assignment.

On the one hand, when $\delta=0$ where we impose homogeneity of CATEs across sites, the rejection rate of the test should approach the nominal significance level (5\%) as $N$ increases, reflecting correct size of the test. On the other hand, when $\delta=1$ where we impose heterogeneous CATEs across sites, the rejection rate should increase with $N$, demonstrating the correct power of the test. We assess the performance of the proposed test using several summary measures. Across $R=1000$ Monte Carlo replications, we report the average estimate ($\hat{\theta}$) based on \eqref{score1}, its standard deviation (std), and the average estimated standard error (mean se). We report empirical rejection rates at the 5\% level, interpreted as the size when $\delta=0$ and power when $\delta\neq 0$, as functions of  $N$ and $\delta$. Finally, we report the effective sample size as a function of the trimming rate.

\begin{table}[H] 
    \input{tables/Table_Sim_Size_delta0_final}    
\end{table}

Table \ref{tab:sim_size_delta0_exp} reports the simulation results based on our main specification when $\delta = 0$ across experimental sites only. That is, when the null hypothesis of homogeneous CATEs across experimental sites is true. The average estimate $\hat\theta$ of the test decreases toward zero as the sample size increases, consistent under the null. The standard deviation and the average standard error decrease by roughly half when the sample size $N$ quadruples, indicating that the estimator is root-$N$ consistent. The empirical rejection rate is slightly above the nominal 5\%. However, it moves towards the correct levels as $N$ increases. Lastly, the effective sample size is close to the nominal sample size $N$ in all specifications, suggesting that trimming is limited and weights are stable. Overall, the results indicate that our test behaves correctly under homogeneous treatment effects across experimental sites and is well calibrated under the null. 

\begin{table}[H]
    \input{tables/Table_Sim_Power_delta1_final}

\end{table}

Table \ref{tab:sim_power_delta1_exp} reports the simulation results based on our main specification when $\delta = 1$, where CATEs vary across experimental sites and the null hypothesis of homogeneous CATEs is false. In this case, the test should reject with high probability. The average estimate $\hat\theta$ becomes increasingly negative as the sample size increases. Both the standard deviation and the mean estimated standard error of the estimator decrease roughly by half when the sample size quadruples, again consistent with root-$N$ convergence. The empirical rejection rate rises sharply with sample size increasing from 79\% to 100\% for larger samples. Finally, the effective sample size remains close to the nominal sample size for all $N$ in all specifications. The results show that our test has strong power to detect violations of effect homogeneity across sites. 

We now consider the setting which combines experimental and observational sites in the absence of confounding where $\rho = 0$. This mixed design mirrors many empirical applications in which treatment is randomized in some sites but not in others, with the latter relying on observational variation. This allows us to examine whether our method can distinguish violations of homogeneity from violations of unconfoundedness. When experimental sites indicate homogeneous CATEs, systematic differences between experimental and observational CATEs provide evidence against the validity of the observational identification strategy.

\begin{table}[H]
    \input{tables/Table_Sim_Size_delta0_obs_exp_noconf}
\end{table}

Table \ref{tab:sim_size_delta0_exp_obs_noconf} reports simulation results for the mixed experimental–observational design under the null hypothesis of homogeneous treatment effects across sites, $\delta = 0$, and no unobserved confounding in the observational sites, $\rho = 0$. Consequently, the identifying assumptions hold, and the test should reject at the nominal significance level. The mean estimate $\hat{\theta}$ moves toward zero as $N$ increases, consistent with the null. The standard deviation and the average estimated standard error decline at the expected rate. The empirical rejection rate is above the nominal 5\% level in smaller samples but it declines with sample size. Finally, the effective sample size remains close to the nominal sample size in all designs. Overall, the results indicate that our test behaves as expected when experimental and observational sites are combined and the observational identifying assumptions hold, although it exhibits some over-rejection in smaller samples.

We next consider a mixed design which combines experimental and observational sites, but now we allow for unobserved confounding in the observational sites captured by $\rho = 0.5$. This value introduces strong confounding in the observational sites. In this setting, the observational identifying assumptions fail, so discrepancies between experimental and observational CATEs are driven by confounding rather than by true effect heterogeneity. As a result, the test may experience distortions in its size, reflecting sensitivity of the test to violations of the identifying assumptions in the observational sites.

\begin{table}[H] 
    \input{tables/Table_Sim_Size_delta0_obs_exp_05_conf}
\end{table}

Table \ref{tab:sim_size_delta0_exp_obs_conf_strong} reports the results for the mixed experimental-observational design under homogeneous treatment effects across sites where $\delta = 0$ and strong unobserved confounding in the observational sites where $\rho = 0.5$. The mean estimate $\hat{\theta}$ decreases at a slower rate than under the scenario of no confounding as the sample size increases. However, both the Monte Carlo standard deviation and the mean estimated standard error shrinks at the expected root-$N$ rate. The effective sample size also remains close to the nominal sample size. Moreover, the rejection rates increase substantially even for moderate sample sizes. The results suggest that, as $N$ grows, our test increasingly detects differences in CATEs between experimental and observational sites which are driven by unmeasured confounding rather than by true treatment effect heterogeneity across sites.

\begin{table}[H] 
    \input{tables/Table_Sim_Power_delta1_obs_exp_noconf}
\end{table}

Table \ref{tab:sim_power_delta1_exp_obs_noconf} reports the results for the mixed experimental-observational design under heterogeneous treatment effects across sites where $\delta = 1$ and there is no unobserved confounding where $\rho = 0$ in the observational sites. Across sample sizes, the mean test statistic $\hat{\theta}$ is negative and becomes slightly more negative as $N$ increases. The standard deviation and the estimated mean standard error decrease approximately at the root rate $N$. Consistent with this, the rejection rate rises as the sample size increases and reaches essentially one for $N \ge 2000$. The effective sample size remains close to the nominal $N$ in all cases. Overall, the simulation results indicate that our test has high power to detect cross-site heterogeneity when identification is valid in both experimental and observational sites and its power increases rapidly with sample size.

\begin{table}[H] 
    \input{tables/Table_Sim_Power_delta1_obs_exp_05_conf}
\end{table}

Finally, Table \ref{tab:sim_power_delta1_exp_obs_conf_strong_partial} reports the results for the mixed experimental-observational design under heterogeneous treatment effects across sites where $\delta = 1$ and unobserved confounding in the observational sites where $\rho = 0.5$. The results show that the test statistic becomes increasingly negative as the sample size increases while the standard deviation and average standard error decrease at root $N$. Regarding power, the results show that the test has lower power in small samples, while the rejection rates increase sharply as the sample size grows. This indicates that strong confounding can weaken the detection of effect heterogeneity in finite samples, but the test regains power for larger sample sizes. Finally, the retained sample size is close to the original sample size. Importantly, because effect heterogeneity and confounding are present simultaneously in this scenario, the test detects differences in treatment effects between experimental and observational sites, but it cannot attribute that difference uniquely to true treatment effect heterogeneity versus violations of the identifying assumptions in the observational sites. Overall, the simulations show that our proposed test is well behaved when its identifying assumptions hold and becomes increasingly informative as the sample size grows.

%% file: tables/Table_Sim_Size_delta0_final.tex
\caption{Simulations: Size under $\delta=0$ across experimental sites.}
\label{tab:sim_size_delta0_exp}
\centering
\begin{tabular}{rrrrrrr}
  \hline
N & \(\hat{\theta}\) & std & mean se & reject 5\% & n\_eff (mean) \\ 
  \hline
500 & 0.041 & 0.063 & 0.063 & 10\% & 492 \\ 
  2000 & 0.027 & 0.028 & 0.028 & 8\% & 1994 \\ 
  8000 & 0.005 & 0.014 & 0.013 & 8\% & 7994 \\ 
   \hline
\end{tabular}
\begin{minipage}{0.95\linewidth}\footnotesize
\textit{Notes.} ‘$N$’ is the sample size per replication. ‘$\hat{\theta}$’ is the average of the test statistic; ‘std’ is the standard deviation; ‘mean se’ is the average estimated standard error. ‘reject 5\%’ is the fraction of replications with $p<0.05$ (empirical size under $\delta=0$, power under $\delta=1$). ‘n\_eff (mean)’ is the average effective sample size under normalized weights. Baseline: $K=5$ folds, Lasso, $L=2$, $p=100$, $\varepsilon=0.05$.
\end{minipage}

%% file: tables/Table_Sim_Power_delta1_final.tex
\caption{Simulations: Power under $\delta=1$ across experimental sites.}
\label{tab:sim_power_delta1_exp}
\centering
\begin{tabular}{rrrrrr}
  \hline
N & \(\hat{\theta}\) & std & mean se & reject 5\% & n\_eff (mean) \\ 
  \hline
500 & -0.204 & 0.073 & 0.073 & 79\% & 492 \\ 
2000 & -0.231 & 0.030 & 0.032 & 100\% & 1993 \\ 
8000 & -0.243 & 0.015 & 0.015 & 100\% & 7994 \\ 
   \hline
\end{tabular}
\begin{minipage}{0.95\linewidth}\footnotesize
\textit{Notes.} ‘$N$’ is the sample size per replication. ‘$\hat{\theta}$’ is the average of the test statistic; ‘std’ is the standard deviation; ‘mean se’ is the average estimated standard error. ‘reject 5\%’ is the fraction of replications with $p<0.05$ (empirical size under $\delta=0$, power under $\delta=1$). ‘n\_eff (mean)’ is the average effective sample size under normalized weights. Baseline: $K=5$ folds, Lasso, $L=2$, $p=100$, $\varepsilon=0.05$.
\end{minipage}

%% file: tables/Table_Sim_Size_delta0_obs_exp_noconf.tex
\caption{Simulations: Size under $\delta=0$ \& $\rho=0$ across experimental \& observational sites.}
\label{tab:sim_size_delta0_exp_obs_noconf}
\centering

\begin{tabular}{rrrrrrr}
  \hline
N & \(\hat{\theta}\) & std & mean se & reject 5\% & n\_eff (mean) \\ 
  \hline
500 & 0.057 & 0.064 & 0.065 & 14\% & 489 \\ 
  2000 & 0.020 & 0.027 & 0.028 & 9.1\% & 1971 \\ 
  8000 & 0.006 & 0.013 & 0.014 & 7.2\% & 7854 \\ 
   \hline
\end{tabular}

\begin{minipage}{0.95\linewidth}\footnotesize
\textit{Notes.} ‘$N$’ is the sample size per replication. ‘$\hat{\theta}$’ is the average of the test statistic; ‘std’ is the standard deviation; ‘mean se’ is the average estimated standard error. ‘reject 5\%’ is the fraction of replications with $p<0.05$ (empirical size under $\delta=0$, power under $\delta=1$). ‘n\_eff (mean)’ is the average effective sample size under normalized weights. Baseline: $K=5$ folds, Lasso, $L=2$, $p=100$, $\varepsilon=0.05$.
\end{minipage}

%% file: tables/Table_Sim_Size_delta0_obs_exp_05_conf.tex
\caption{Simulations: Size under $\delta=0$ \& $\rho=0.5$ across experimental \& observational sites.}
\label{tab:sim_size_delta0_exp_obs_conf_strong}
\centering

\begin{tabular}{rrrrrr}
  \hline
N & \(\hat{\theta}\) & std & mean se & reject 5\% & n\_eff (mean) \\ 
  \hline
500  & 0.174 & 0.070 & 0.068 & 72.6\% & 489 \\ 
2000 & 0.137 & 0.028 & 0.030 & 99.8\% & 1974 \\ 
8000 & 0.123 & 0.014 & 0.014 & 100\%  & 7870 \\ 
   \hline
\end{tabular}

\begin{minipage}{0.95\linewidth}\footnotesize
\textit{Notes.} ‘$N$’ is the sample size per replication. ‘$\hat{\theta}$’ is the average of the test statistic; ‘std’ is the standard deviation; ‘mean se’ is the average estimated standard error. ‘reject 5\%’ is the fraction of replications with $p<0.05$ (empirical size under $\delta=0$, power under $\delta=1$). ‘n\_eff (mean)’ is the average effective sample size under normalized weights. Baseline: $K=5$ folds, Lasso, $L=2$, $p=100$, $\varepsilon=0.05$.
\end{minipage}

%% file: tables/Table_Sim_Power_delta1_obs_exp_noconf.tex
\caption{Simulations: Power under $\delta=1$  \& $\rho=0$ across experimental \& observational sites.}
\label{tab:sim_power_delta1_exp_obs_noconf}
\centering
\begin{tabular}{rrrrrr}
  \hline
N & \(\hat{\theta}\) & std & mean se & reject 5\%  & n\_eff (mean) \\ 
  \hline
500 & -0.177 & 0.073 & 0.074 & 67.7\%  & 488 \\ 
  2000 & -0.225 & 0.031 & 0.032 & 100\%  & 1972 \\ 
  8000 & -0.242 & 0.015 & 0.015 & 100\%  & 7855 \\ 
   \hline
\end{tabular}

\begin{minipage}{0.95\linewidth}\footnotesize
\textit{Notes.} ‘$N$’ is the sample size per replication. ‘$\hat{\theta}$’ is the average of the test statistic; ‘std’ is the standard deviation; ‘mean se’ is the average estimated standard error. ‘reject 5\%’ is the fraction of replications with $p<0.05$ (empirical size under $\delta=0$, power under $\delta=1$). ‘n\_eff (mean)’ is the average effective sample size under normalized weights. Baseline: $K=5$ folds, Lasso, $L=2$, $p=100$, $\varepsilon=0.05$.
\end{minipage}

%% file: tables/Table_Sim_Power_delta1_obs_exp_05_conf.tex
\caption{Simulations: Power under $\delta=1$ \& $\rho=0.5$ across experimental \& observational sites.}
\label{tab:sim_power_delta1_exp_obs_conf_strong_partial}
\centering

\begin{tabular}{rrrrrr}
  \hline
N & \(\hat{\theta}\) & std & mean se & reject 5\% & n\_eff (mean) \\ 
  \hline
500  & -0.063 & 0.080 & 0.078 & 15.1\% & 489 \\ 
2000 & -0.108 & 0.033 & 0.034 & 89.8\% & 1975 \\ 
8000 & -0.125 & 0.016 & 0.016 & 100\% & 7869 \\ 
  \hline
\end{tabular}

\begin{minipage}{0.95\linewidth}\footnotesize
\textit{Notes.} ‘$N$’ is the sample size per replication. ‘$\hat{\theta}$’ is the average of the test statistic; ‘std’ is the standard deviation; ‘mean se’ is the average estimated standard error. ‘reject 5\%’ is the fraction of replications with $p<0.05$ (empirical size under $\delta=0$, power under $\delta=1$). ‘n\_eff (mean)’ is the average effective sample size under normalized weights. Baseline: $K=5$ folds, Lasso, $L=2$, $p=100$, $\varepsilon=0.05$.
\end{minipage}

%% file: data.tex
\section{Application} \label{sec:data}

We illustrate our method using data from the International Stroke Trial (IST), a large multi-centre randomized controlled trial in acute ischaemic stroke conducted by the IST Collaborative Group \citep{Sandercock2011}. The IST investigated whether early administration of aspirin, heparin, both, or neither affects clinical outcomes after stroke. Patients were eligible if they had a clinical diagnosis of acute ischaemic stroke within 48 hours of symptom onset and had no clear indication for, or contraindication to, either treatment. After a CT scan to support the diagnosis, clinicians contacted a central randomization service that recorded baseline characteristics and returned the assigned treatment.

The dataset contains anonymized individual-level information on 19{,}435 patients treated in 467 hospitals across 36 countries. It includes baseline characteristics, clinical status at randomization, short-run outcomes measured at 14 days, and follow-up outcomes at six months. The primary outcome of interest of the trial was death or dependency in daily living six months after randomization. Our empirical analysis focuses on the randomized assignment to aspirin. We define the treatment indicator $D$ as assignment to aspirin and the outcome $Y$ as an indicator equal to one if the patient is dead or dependent at six months, and zero otherwise. We restrict the sample to patients with non-missing information on treatment assignment and the six-month outcome. Additionally, for comparability, we focus on observations from the main trial and discard observations from the pilot trial. 

\begin{table}[H]
\caption{Sample construction.}
\centering
\begin{tabular}{lr}
\hline
Step & $N$ \\
\hline
Full IST dataset & 19{,}435 \\
Main trial only (drop pilot) & 18{,}451 \\
Non-missing $D$ and $Y$ & 18{,}273 \\
Final analysis sample ($\ge 50$ per country) & 18{,}189 \\
\hline
\end{tabular}
\label{tab:ist_sample_flow}
\end{table}

Table \ref{tab:ist_sample_flow} summarizes the sample construction. Starting from the full IST dataset, we first observations from the pilot phase to focus on the main trial for comparability reasons. We then restrict the sample to patients with non-missing treatment assignment and six-month outcome. Finally, we impose a minimum sample size requirement at the site level and retain only countries with at least 50 observation. The final analysis sample contains $N=18{,}189$ patients from 31 countries listed in Table \ref{tab:sample_country}. As baseline covariates $X$, we use pre-treatment patient characteristics measured at randomization: age, sex, systolic blood pressure, indicators for baseline level of consciousness (fully alert, drowsy, unconscious), and whether a CT scan was performed before randomization. These covariates capture clinically relevant differences at baseline health between patients.

\begin{table}[H]
\centering
\caption{Baseline covariate balance: Standardized Differences in Means.}
\label{tab:baseline_SMD}
\begin{tabular}{lccc}
\hline
\textbf{Variable} & \textbf{Mean Control (SD)} & \textbf{Mean Treated (SD)} & \textbf{SMD} \\
\hline
Age & 71.87 (11.53) & 71.89 (11.61) & 0.002 \\
Systolic blood pressure & 160.45 (27.63) & 160.04 (27.83) & 0.015 \\
Female & 0.46 (0.50) & 0.47 (0.50) & 0.018 \\
CT before randomization & 0.68 (0.47) & 0.67 (0.47) & 0.016 \\
Fully alert  & 0.77 (0.42) & 0.77 (0.42) & 0.003 \\
Drowsy & 0.22 (0.41) & 0.22 (0.41) & 0.003 \\
Unconscious & 0.01 (0.12) & 0.01 (0.12) & $<0.001$ \\
$N$ & 9{,}101 & 9{,}088 & \\
\hline
\end{tabular}

\begin{minipage}{0.95\linewidth}\footnotesize
\textit{Notes.} Entries report mean (standard deviation). SMD denotes the standardized mean difference between treated and control groups.
\end{minipage}
\end{table}

\begin{figure}[H]
    \centering
    \caption{Outcome rates by country (site): Pr(dead or dependent at 6 months) with 95\% confidence intervals.}
    \includegraphics[width=0.7\linewidth]{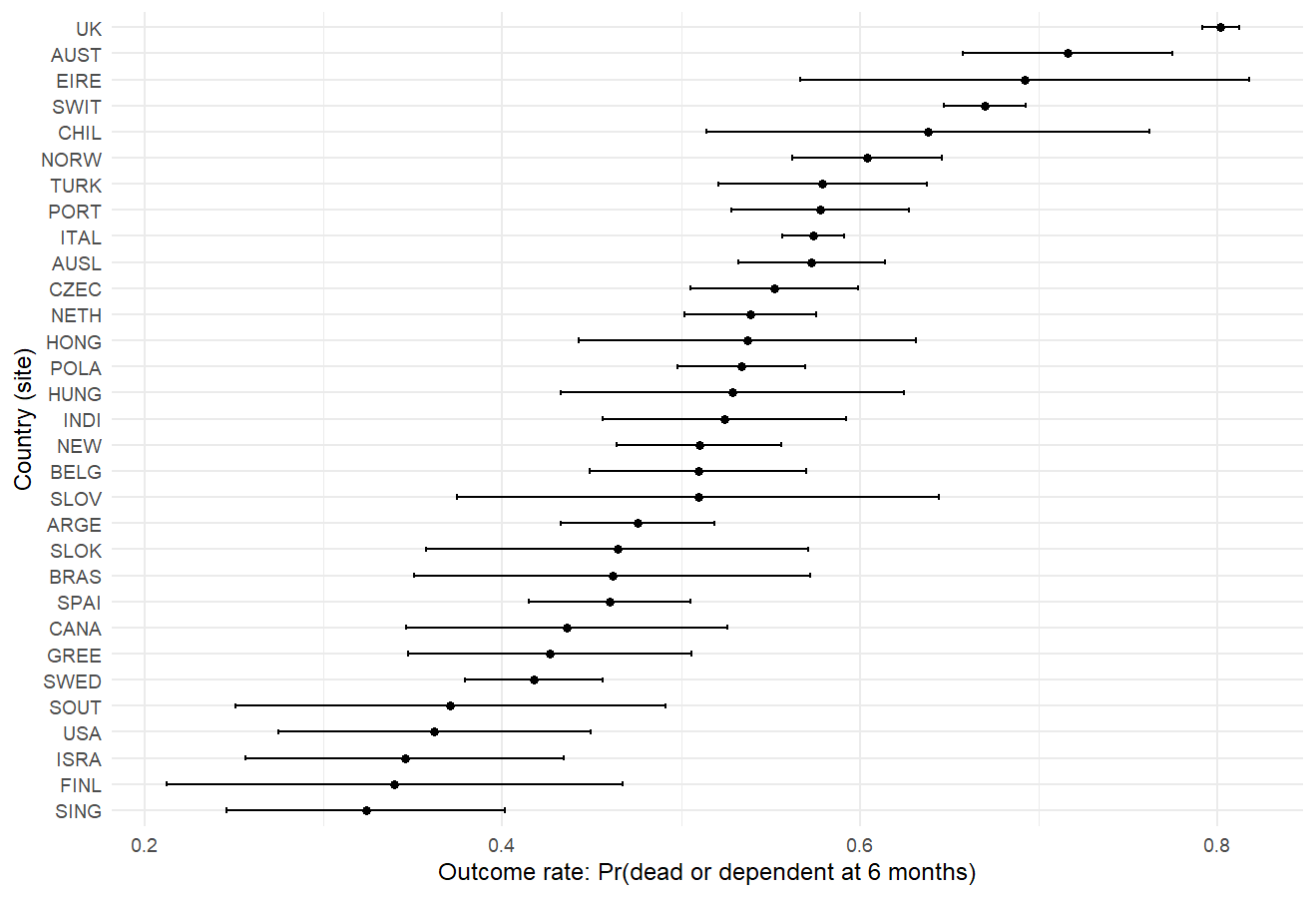}
    \label{fig:sites_outcome_rates}
\end{figure}

Table \ref{tab:baseline_SMD} summarizes baseline covariate balance between treated and control patients using standardized difference in means (SMDs). The treated and control groups are very similar across all characteristics as all SMDs are below conventional thresholds for meaningful imbalance. In particular, age, systolic blood pressure, sex, pre-randomization CT use, and baseline consciousness status are nearly identical across treatment arms. These patterns support a successful treatment randomization in the IST study. In addition, treatment assignment is also balanced within each country, although sample sizes within countries vary substantially (see Figure \ref{fig:sites_shares} and Table \ref{tab:sample_country}).
In this context, we take countries as experimental sites, indexed by $Z$, and apply our test of effect homogeneity between these countries. This setting is well suited to our approach because a common randomized protocol was implemented across all participating countries. However, clinical environments and baseline risk may differ between sites, as can be seen in Figure \ref{fig:sites_outcome_rates}. 

Figure \ref{fig:sites_outcome_rates} displays outcome rates by country. The outcome rates vary markedly across sites, ranging from 0.324 to 0.802, which likely reflects differences in baseline risk, case mix, and clinical practice across countries. This heterogeneity in outcome levels motivates testing whether the treatment effects are homogeneous across countries, specifically if the conditional effect of aspirin on death or dependency is stable between countries. We therefore apply our proposed test of CATE homogeneity across multiple experimental sites to the IST dataset. The estimand is the average treatment effect of randomized aspirin assignment on six-month death or dependency, adjusting for baseline covariates. In the spirit of our framework, failing to reject homogeneity indicates no systematic interactions between treatment and unobserved determinants of outcomes in the mean effect and supports generalizing the estimated effect across countries. In contrast, rejecting homogeneity suggests that the effect varies across sites in ways not captured by observed covariates, consistent with treatment–unobservable interactions and/or other forms of site-level heterogeneity.

\begin{table}[H]
\caption{Test of effect homogeneity across experimental sites (countries)}
\centering
\begin{tabular}{lrrrr}
\hline
$\varepsilon$ & $\hat{\theta}$ & se & p-value & $n_{\text{eff}}$ \\
\hline
0.05  & 0.012 & 0.006 & 0.046 & 15805 \\
0.10  & 0.002 & 0.004 & 0.706 & 9349  \\
\hline
\end{tabular}

\begin{minipage}{0.95\linewidth}\footnotesize
\textit{Notes.} $\varepsilon$ is the trimming threshold for the propensity-score components used to construct weights. $n_{\text{eff}}$ is the effective sample size after trimming.
\end{minipage}
\label{tab:ist_test_main}
\end{table}

Table \ref{tab:ist_test_main} reports the results of our test of CATE homogeneity across countries, implemented with two trimming thresholds, $\varepsilon\in{0.05,0.10}$. For the baseline threshold $\varepsilon=0.05$, the test rejects homogeneity at the 5\% level ($p=0.046$), providing marginal evidence that the conditional effect of aspirin may vary across countries. Increasing trimming to $\varepsilon=0.10$ yields a much smaller test statistic and a large $p$-value ($p=0.706$), so we no longer reject homogeneity. The shift in inference is accompanied by a sharp decline in the effective sample size. 

The results show the impact that trimming and therefore, overlap can have on the conclusions. Importantly, in this context, the sensitivity of the results to trimming is not due to the limited overlap in treatment assignment, as aspirin is randomized within each country and the estimated propensity scores are tightly concentrated around $0.5$. Rather, it reflects limited support for certain country covariate combinations. In some countries, specific covariate profiles are rare and/or the country sample size is small, resulting in estimated site-specific propensity scores that can be very low. This poses a problem because the estimation weights are based on the inverse of the propensity scores, so small scores translate into very large weights. As a result, a small number of observations in sparsely supported regions of the country-specific covariate distribution can receive extreme weights and disproportionately influence the test statistic. Trimming prevents this by excluding such cases with extreme propensity scores. With stricter trimming that focuses on regions with improved common support, the test does not reject the null hypothesis of homogeneous treatment effects across countries.

%% file: discussion.tex
\section{Conclusion} \label{sec:discussion}

In this work, we introduced a framework for testing the homogeneity of conditional average treatment effects (CATEs) across multiple experimental and observational sites. The proposed test is built on a Neyman orthogonal score that extends \citet{apfel2024learningcontrolvariablesinstruments} to a double difference setting. Under specific regularity conditions (in particular, $o(n^{-1/4})$ convergence rates for the nuisance estimators) and with cross-fitting, the resulting estimator is $\sqrt{n}$-consistent and asymptotically normal. We also showed how the same logic carried over to settings with alternative identification strategies, such as instrumental variables and panel designs with parallel trends. 

The simulation study indicated that the test is well behaved when the identifying assumptions hold and becomes increasingly informative as sample size grows. In randomized multi-site designs, the test has the correct size under homogeneity and has high power against heterogeneity. In mixed designs that combine experimental and observational sites, the test rejects systematically when unobserved confounding is present in the observational sites, even when treatment effects are homogeneous, highlighting the usefulness of the test as a diagnostic for flagging potential confounding. 

We then illustrated the approach using data from the International Stroke Trial, treating countries as sites and testing whether the conditional effect of randomized aspirin assignment on six month death or dependency is homogeneous across countries.  With more permissive trimming, we reject homogeneity, while with stricter trimming we do not reject. This sensitivity likely arises because some countries have very few observations for certain patient profiles. With a more permissive trimming rule, these rare profiles can receive very large weights and therefore have a disproportionate impact on the results. However, with stricter trimming, we exclude those poorly represented cases, and the analysis relies on patient profiles that are more common within each country.

Overall, the proposed framework provides a practical and flexible tool for assessing homogeneity of CATEs across experimental and observational data. The test helps researchers to diagnose confounding as well as to evaluate the internal and external validity of their estimates. This is increasingly valuable as researchers now often have access to both randomized trials and rich observational datasets, but lack principled ways to determine when estimates from these sources can be compared, combined, or extrapolated across settings.